\documentclass[journal=apchd5,manuscript=article]{achemso}

\usepackage[version=3]{mhchem} 
\usepackage[T1]{fontenc}       
\usepackage[numbers,sort&compress]{natbib}
\usepackage{graphicx}

\author{Yuting Guo$^\S$}
\affiliation{Department of Engineering, University of Cambridge, Cambridge CB3 0FA, United Kingdom}
\author{Jonathan Bar-David$^\S$}
\affiliation{Department of Engineering, University of Cambridge, Cambridge CB3 0FA, United Kingdom}
\author{Pasquale Cilibrizzi}
\affiliation{Department of Engineering, University of Cambridge, Cambridge CB3 0FA, United Kingdom}
\author{Sung-Yul L. Park}
\affiliation{Center for Opto-Electronic Materials and Devices Research, Korea Institute of Science and Technology, Seoul, South Korea}
\author{Jin Dong Song}
\affiliation{Center for Opto-Electronic Materials and Devices Research, Korea Institute of Science and Technology, Seoul, South Korea}
\author{Luca Sapienza}

\affiliation{Department of Engineering, University of Cambridge, Cambridge CB3 0FA, United Kingdom}

\email{LS2052@cam.ac.uk}

\title[An \textsf{achemso} demo]
  {Thin-capping layer epitaxial quantum dots for near-field quantum photonics} 

\begin{document}

\let\thefootnote\relax\footnotetext{$^\S$Equal contribution}

\begin{abstract}
Epitaxial quantum dots (QDs) are widely recognised as one of the best quantum light sources, given their good stability, brightness, quantum efficiency and coherence. To reach such properties, QDs are protected from potentially detrimental surface states by a relatively thick capping layer, typically exceeding 50\,nm. This prevents the implementation of near-field effects, like plasmonic-based ones, that can dramatically increase the light-matter interaction, since the control of the spontaneous emission dynamics and directionality of the emission can only occur if the emitter is in close proximity (typically tens of nanometers or less) from the metallic structures.  In this work, we report the growth and optical characterisation of InAs/GaAs QDs with GaAs capping layer thickness of 10\,nm, 20\,nm, and a more standard 95\,nm. Remarkably, we observe emission linewidths up to 5 times smaller than previously reported, with unperturbed excitonic lifetimes. These results demonstrate that the epitaxial QD's high optical quality can be maintained even when the emitters are at reduced distances from the surface, opening the path for the exploration of near-field light–matter interactions with coherent and stable emitters, suitable for quantum technology applications.
\end{abstract}

\section*{Keywords}
Quantum photonics, Epitaxial quantum dots, Plasmonics, Near-field effects 

\maketitle
\vspace{1cm}
Epitaxial QDs have shown to be mature quantum light emitters for fundamental quantum optics studies and for quantum technology applications \cite{SPS_review}. In particular, InAs/GaAs QDs can shown desirable properties, like high brightness \cite{nanowire}, high single-photon purity \cite{Sapienza_QD_2015}, and excellent indistinguishability, reaching transform-limited linewidths \cite{Warburton}. In order to improve the properties of QDs, these emitters are often coupled to dielectric photonic structures designed to guide the emission into a desired optical mode \cite{cQED} and to control the spontaneous emission rate via the Purcell effect \cite{cQED, Purcell}. 

Another class of photonic devices that allow strong increases of the light-matter interaction are plasmonic nanoantennae, based on near-field effects at metal-dielectric interfaces \cite{Plasmonics1, Plasmonics2, Plasmonics3, JJB}. However, such devices require close proximity (typically 10\,nm or less), in order for the emitters to couple to the electromagnetic near-field confined mode. Typically, dye molecules \cite{molec_plasmonics} or colloidal QDs \cite{JJB_QDs, Strauf_cube} are therefore utilised: however, such emitters can suffer from linewidth broadening \cite{Nelson_CQD_2024}, spectral diffusion \cite{Beyler_CQD_2013}, blinking \cite{Nirmal_blinking_1996} and photobleaching during prolonged excitation \cite{Sark_Bleaching_2002}. Such limited stability, therefore prevents their suitability for scalable quantum technology applications. Epitaxially-grown droplet GaAs QDs have been investigated as a possible alternative \cite{our_QDs}, however they are limited by emission wavelengths in the 700\,nm range. Otherwise, Stranski-Krastanov \cite{SK_QD} InAs/GaAs QDs have been coupled to metallic rings to increase the light collection efficiency \cite{our_rings1, our_rings2, our_rings3, our_rings4}, but such devices showed no changes in emission recombination dynamics, and the overall performances were limited by the 95\,nm distance between the emitter and the metal/dielectric interfaces where the electromagnetic field is confined.

\begin{figure*}[t!]
\centering
\includegraphics[width=1\linewidth]{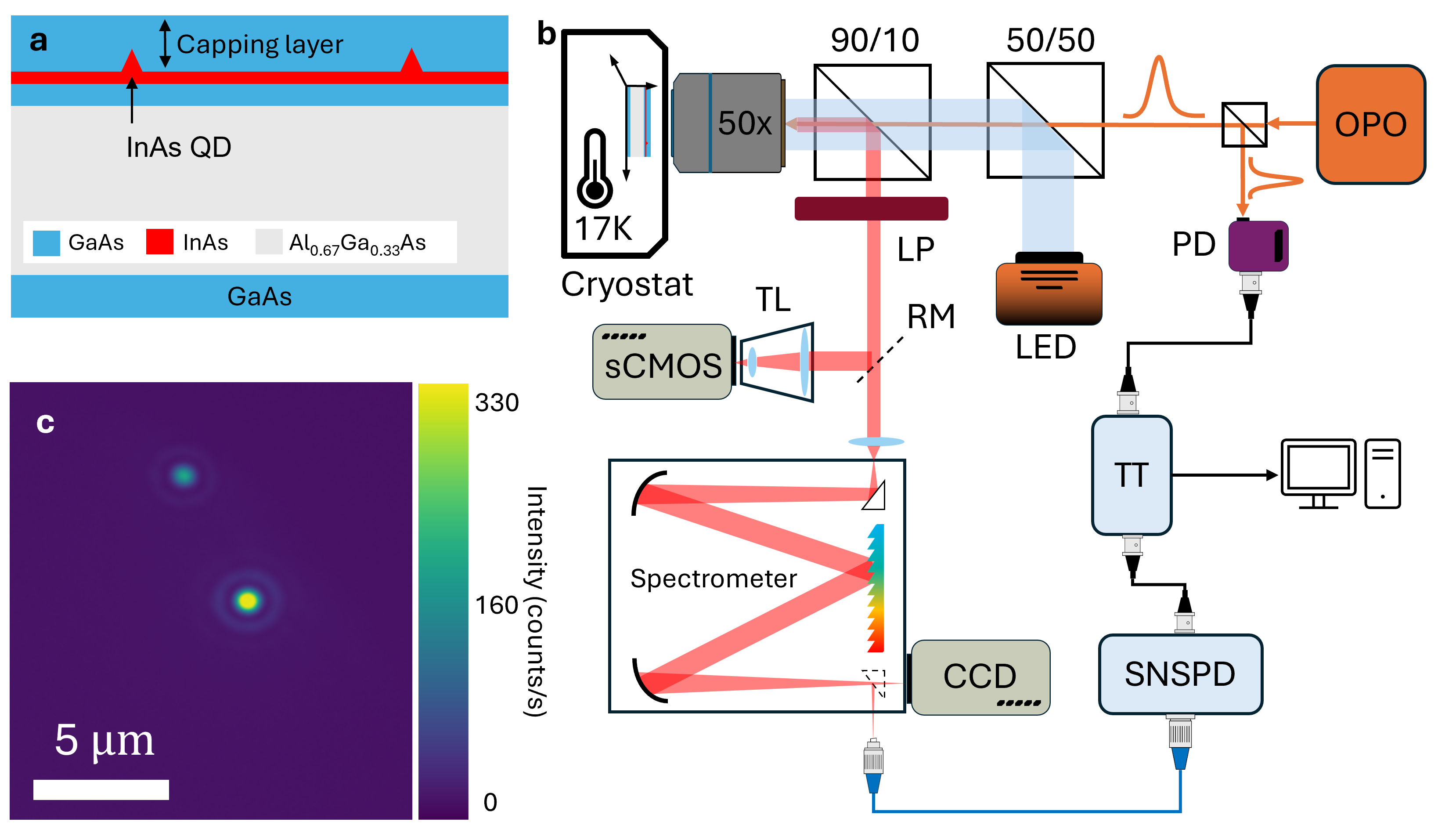}
\caption{(a) Schematic of the sample (not to scale). The triangles represent the QD single-photon emitters. The capping layer thickness is highlighted. (b) Schematic of the experimental setup (not to scale). The sample is illuminated through a 50$\times$ objective by an LED for imaging and a pulsed laser for spectroscopy and time-resolved measurements. A 90/10 (Reflection/Transmission) beamsplitter reflects the photoluminescence signal towards the collection path. For imaging, photoluminescence is sent onto a scientific Complementary Metal–Oxide–Semiconductor (sCMOS) camera via a tube lens (TL). For spectroscopy, light is dispersed by a spectrometer and recorded by a Charge-Coupled Device (CCD). For time-resolved measurements, light is coupled to a single-mode fibre leading to superconducting nanowire single-photon detectors (SNSPDs) connected to a time-tagger (TT), triggered by a fast photo-diode (PD). LP\,=\,900\,nm long-pass filter, RM\,=\,removable mirror. (c) Photoluminescence image of isolated QDs, collected under 450\,nm-LED illumination.}
\label{fig:exp_setup}
\end{figure*}

The optical properties of QDs are sensitive to detrimental surface-induced effects \cite{Ferdos_QD_2002,houel_probing_2012,zhao_deterministic_2025}: reducing the capping layer thickness can activate non-radiative recombination pathways, and/or introduce electromagnetic noise from edge states and dangling bonds, leading to linewidth broadening and shortened excitonic lifetimes \cite{Michler_QD_2024, wang_optical_2004, heitz_energy_1997, liu_single_2018}. Following previous literature\cite{wang_optical_2004}, to mitigate these effects, a relatively thick GaAs capping layer, typically exceeding 50\,nm, is employed to spatially separate the InAs/GaAs QDs from the surface and preserve their optical quality.

While this approach is effective in maintaining emitter performances, it restricts near-field coupling between the QD and surface photonic structures, like plasmonic resonators \cite{Plasmonics1, Plasmonics2, Plasmonics3, JJB}. As a result, the efficiency of photon extraction and light–matter interaction is inherently limited, and therefore dielectric waveguides and cavities have been chosen instead \cite{QT_review}, taking advantage of more extended optical modes confined by refractive index contrast, rather than metal-dielectric interface, near-field effects. 

In this work, we report on the growth and optical characterisation of InAs/GaAs epitaxial QDs with reduced GaAs capping layer (CL) thickness of 10\,nm and 20\,nm. Despite the proximity to the surface, the measured emission linewidths are found to be narrower than values reported and extrapolated from previous studies \cite{wang_optical_2004, chellu_gaas_2021}, and, as opposed to previous reports \cite{wang_optical_2004}, the excitonic lifetimes remain comparable to those observed in structures with substantially thicker capping layers (95\,nm). These results indicate that good optical quality can be preserved even in thin-capped structures, thus providing a promising platform for exploring Purcell enhancement via plasmonic structures, and improved light guidance via near-field photonic effects.


The samples investigated in this work were grown by molecular beam epitaxy (MBE), using the Stranski-Krastanov technique \cite{SK_QD}: a 2-inch semi-insulating (001) GaAs substrate was loaded onto the substrate manipulator, and the native surface oxide was thermally desorbed at 600$^\circ$C under an As$_{4}$ overpressure. The substrate temperature was then reduced to 580$^\circ$C, and an approximately 200\,nm-thick GaAs buffer layer was grown. Subsequently, a 1\,$\mu$m-thick Al$_{0.67}$Ga$_{0.33}$As layer and a 5\,nm-thick GaAs layer were grown at 590$^\circ$C. The substrate temperature was then lowered to 485$^\circ$C, before growing the InAs QDs.

The InAs deposition corresponded to approximately 1.5 monolayers. Atomic force microscopy measurements of uncapped reference samples showed a low QD density of approximately one QD per 1$\mu$m$\times$1$\mu$m area, with QD heights of up to approximately 4\,nm.

Following QD formation, the QDs were initially capped with 5\,nm of GaAs grown at 485$^\circ$C. The growth was then interrupted, and the substrate temperature was increased to 580$^\circ$C under an As$_{4}$ overpressure. Following this thermal treatment, an additional 5 or 15\,nm of GaAs was grown, resulting in nominal total GaAs capping-layer thicknesses of 10 and 20\,nm, respectively. The sample layer structure is schematically shown in Figure 1(a).

The samples are characterised using a home-built micro-photoluminescence experimental setup schematically shown in Figure 1(b), where the sample, cooled down to $\sim$17\,K, is placed onto a piezoelectric stack that enables movements in three orthogonal directions. 

A 450\,nm Light Emitting Diode (LED) is used to excite the sample for photoluminescence (PL) imaging \cite{Sapienza_QD_2015}, using a 50$\times$ magnification, Numerical Aperture (NA)\,=\,0.75 objective to focus the excitation light and collect the emission. Figure 1(c) shows an example of a PL image of two isolated QDs that can be then selectively excited using a laser with excitation spot of $\mu$m size.

An 880\,nm, 12\,ps-pulsed laser, with 80\,MHz repetition rate is used to excite the sample to spectrally characterise the photoluminescence from single quantum dots. Power series measurements, collected as a function of increasing laser excitation power, are performed to evaluate the emission saturation level and linewidths (see Figure 2), which are critical parameters for assessing QD emission quality. \cite{Michler_QD_2024}. 

\begin{figure*}[t!]
\centering
\includegraphics[width=1\linewidth]{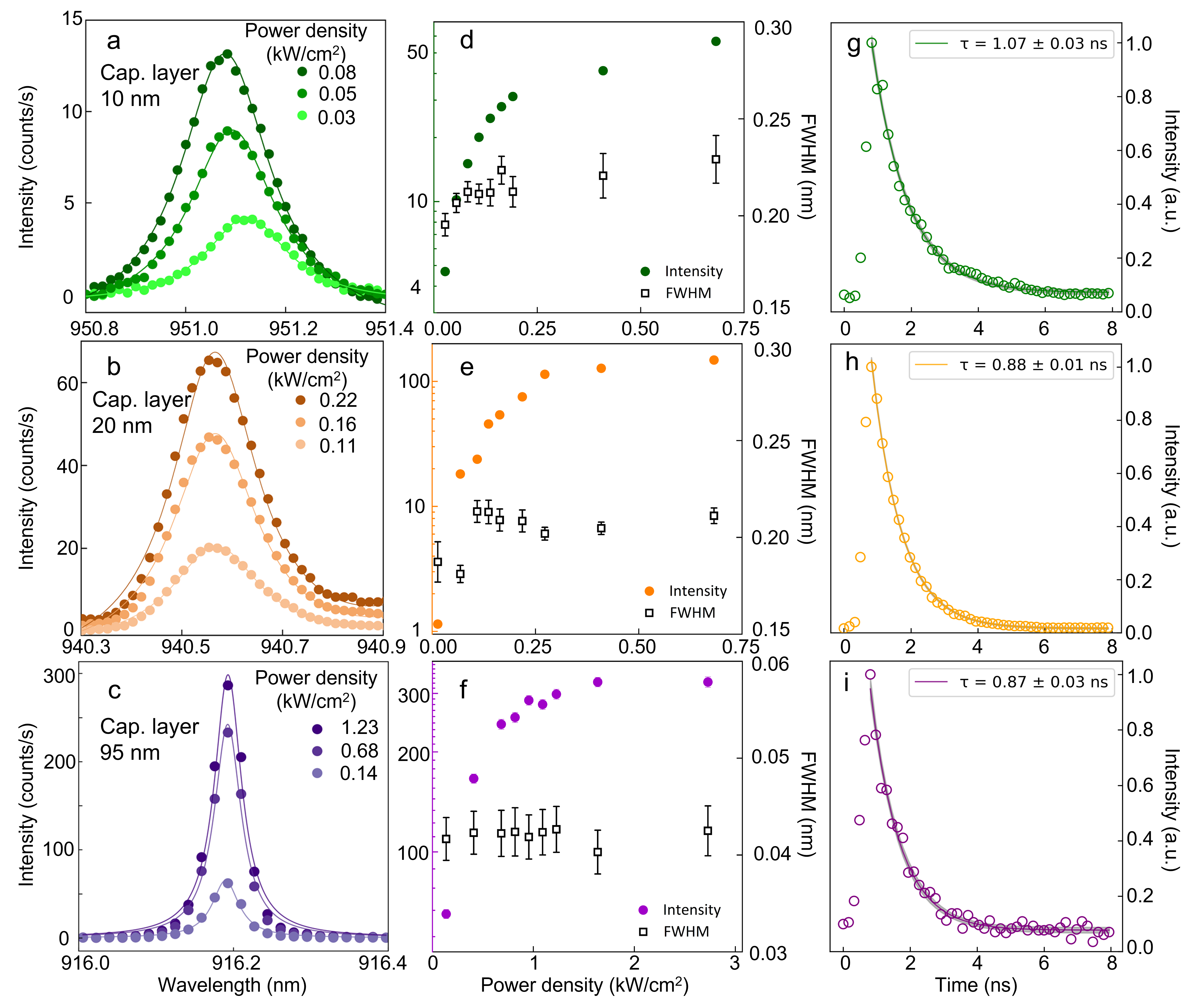}
\caption{(a-c) Photoluminescence emission peaks of InAs/GaAs quantum dots (solid circles, 10\,nm CL in green, 20\,nm CL in orange, 95\,nm CL in purple) taken at three different 880\,nm laser pump powers, fitted (solid lines) with Lorentzian functions. (d-f) Peak intensity (solid circles) and Full Width Half Maximum (FWHM, black hollow squares) of selected QD lines, plotted as a function of laser excitation power density. (g-i) Time-resolved PL measurements, showing the excited-state decays (empty circles) fitted by an exponential decay curve (solid lines). The lifetime extracted from the fit ($\tau$) is shown in the legend and the shaded grey areas correspond to the 95\,$\%$ confidence intervals in the fit.}
\label{raw_results}
\end{figure*}

For time-resolved photoluminescence experiments, the photoluminescence (PL) from the sample is spectrally filtered by a 0.75\,m spectrometer equipped with a 1200 lines/mm grating that allows isolating specific emission lines. These are then coupled into a single-mode fibre leading to superconducting nanowire single-photon detectors, connected to a time-tagger.

Figure 2 shows examples of PL spectra of InAs/GaAs QDs from samples with varying capping layer (CL) thickness. Panels (a-c) show the evolution of a single InAs QD line while increasing the excitation power. For the thin-capped samples, saturation onsets at an excitation power density of $\sim$0.25\,kW/cm$^2$, compared to $\sim$1\,kW/cm$^2$ for the 95\,nm-capped sample, see Fig.2 (d-f). This could be explained by the proximity of the emitter to the surface and therefore a more efficient coupling of the laser power to the QD. It is worth noting that a reduced excitation power is desirable for lower-consumption devices and to reduce thermally-activated mechanisms like phonon and charge capture/release that can reduce quantum light coherence.\\
Figures 2\,(g-i) present the exciton lifetimes measured on the same emission line shown in panels (a-c).  The exciton lifetimes are consistent across all samples, and are fitted with an exponential decay which gives lifetimes in the range 0.9-1.1\,ns, in accordance with the typical 1\,ns lifetime reported for InAs/GaAs quantum dots at cryogenic temperatures \cite{Michler_QD_2024}.
It is worth noting that previous reports \cite{wang_optical_2004} showed a significant decrease in the lifetime when the capping layer thickness was reduced, reaching values of 550\,ps for 20\,nm capped QDs and 65\,ps for 10\,nm capped QDs. Such behaviour was attributed to non-radiative recombination mechanisms taking place when the QD was closer to the surface. The fact that the lifetime of the QD under study are $\sim$1\,ns instead, shows that our emitters are more stable and preserve good optical properties.

\begin{figure}[t!]
\includegraphics[width=\columnwidth, scale=0.8]{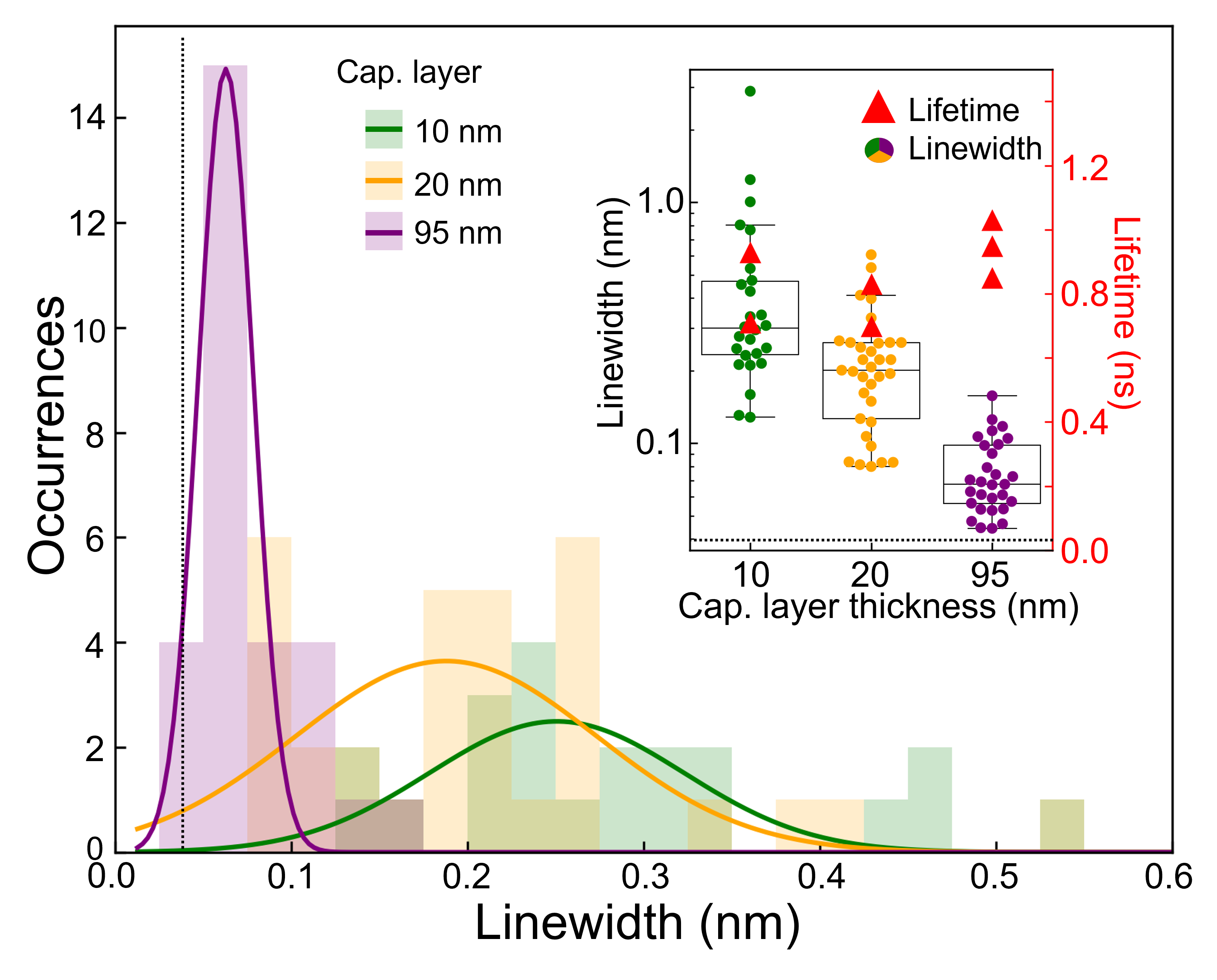}
\caption{Histogram of the linewidths of QDs as a function of capping layer (CL) thickness, with the same colour coding as in Fig.2. Inset: distribution of the linewidths as a function of CL thickness. Black boxes indicate the central 50\,\% of the distribution, with the centre line indicating the median. Whiskers indicate the full data range, excluding outliers. Extracted lifetime values for different capping layer thickness are shown in red triangles. In both the main panel and the inset, a fine dotted line indicates the system resolution limit.}
\label{Distribution}
\end{figure}

In Figure \ref{Distribution}, we present histograms of the entire linewidth population, alongside a Gaussian fit to the distribution. For the samples with CL\,=\,95\,nm (purple), the linewidth distribution is narrower and centred around 0.06\,nm, while for CL\,=\,20 and 10\,nm (orange and green, respectively) the distributions are broader, with interquartile ranges of $\sim$0.14 and $\sim$0.24\,nm, respectively, compared to $\sim$0.04 nm for the 95\,nm capped QDs, indicating a larger variability in the population. Despite these broader distribution, the Gaussian fits show that the linewidth of the QDs with CL\,=\,10\,nm is only broadened by less than 0.1\,nm more than the CL\,=\,20\,nm. In the inset to Fig.3, we present the distribution of the entire linewidth population for the different samples, alongside the lifetime values measured for each sample type, showing lifetimes that remain within the 0.6-1.1\,ns interval.

\begin{figure}[t!]
\includegraphics[width=0.8\columnwidth, scale=0.1]{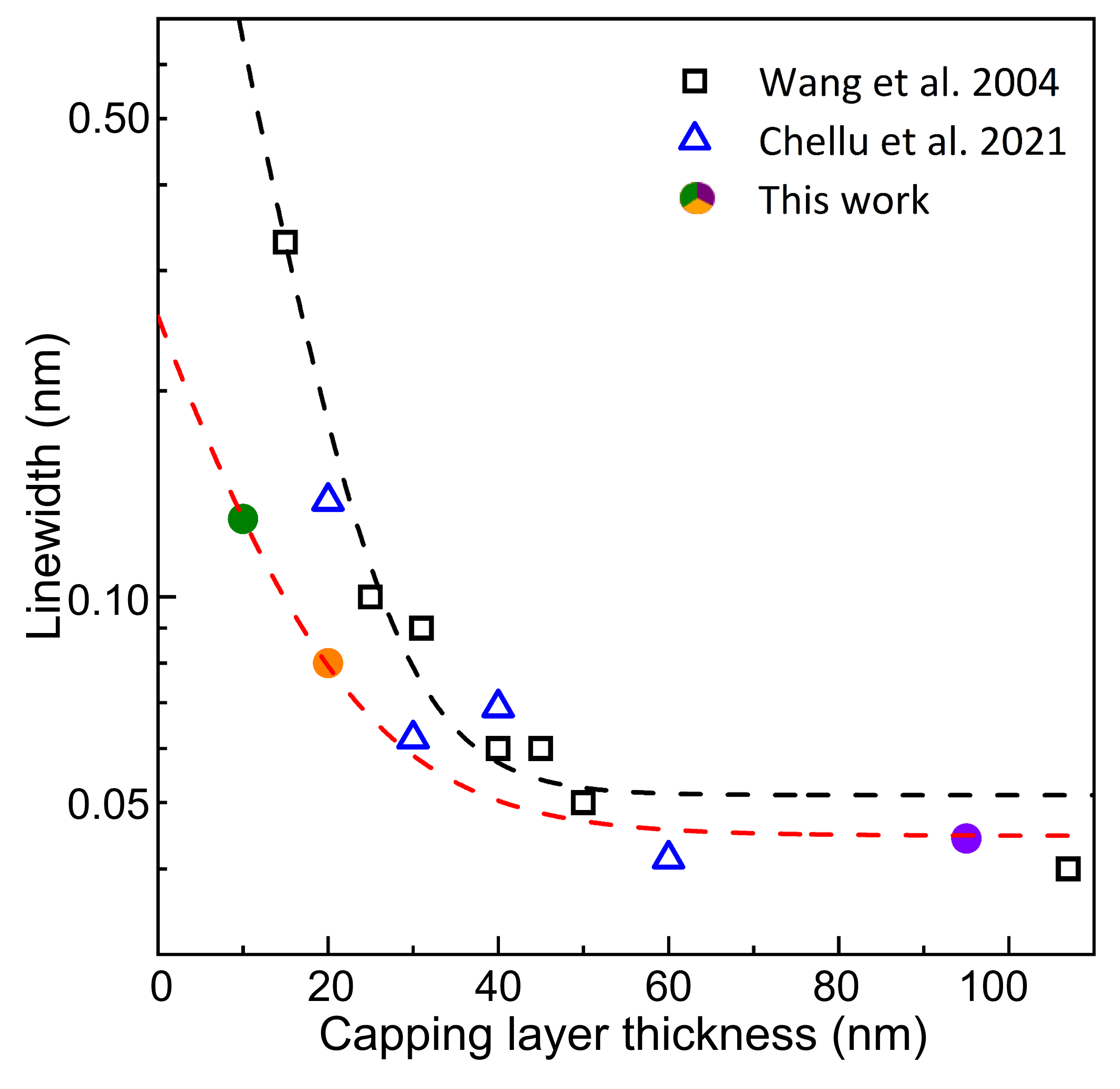}
\caption{Collection of InAs/GaAs QD narrowest emission linewidths, plotted as a function of capping layer thickness (colored solid circles), compared with linewidth values from previous reports \cite{chellu_gaas_2021,wang_optical_2004}. The black dashed line is an exponential fit to Wang et al.'s data \cite{wang_optical_2004}. The red dashed line shows the exponential fit to our data.}
\label{Literature_review}
\end{figure}

In Fig.4, the emission linewidths are compared to those reported in previous studies: by comparing the spectra from sample with different CL thickness, we see that the emission linewidth narrows as the CL thickness increases from our smallest value measured of 0.13\,nm for CL\,=\,10\,nm to 0.08\,nm for CL\,=\,20\,nm, and 0.04\,nm (reaching the spectral resolution limit of our system) for the reference CL\,=\,95\,nm. Spectral broadening can be attributed to spectral wandering due to the presence of charged defects or surface states \cite{spectral_wander}, as previously shown \cite{wang_optical_2004}. However, previous works reported linewidths of 0.14\,nm and 0.18\,nm (extrapolated) for CL\,=\,20\,nm \cite{chellu_gaas_2021}, and an extrapolated linewidth of 0.65\,nm for CL\,=\,10\,nm \cite{wang_optical_2004}: such values are 175\,$\%$ and 500\,$\%$ of our measured values, respectively, showing that our QDs experience reduced dephasing processes compared to those previous reports.As discussed, our spectra are measured under non-resonant excitation, condition where linewidths are larger than transform-limited values \cite{Michler_QD_2024}. However, the measured values are an indication of the level of dephasing mechanisms taking place and making the linewidths depart from the resolution limited values measured for our reference sample (with 95\,nm capping layer thickness).

The preservation of the optical properties observed in our thin-capped QDs may be related to the growth procedure employed before completion of the GaAs capping layer. The strain field associated with InAs QDs can extend considerably beyond the physical dimensions of the QDs themselves: for vertically stacked InAs QDs, interactions between QD layers have been observed with GaAs spacer thicknesses of approximately 20\,nm \cite{Ref1}, while structural correlations between vertically separated QDs have been reported even for spacer thicknesses of approximately 35\,nm \cite{Ref2}. Therefore, although the low-density InAs QDs investigated here have heights of only up to approximately 4\,nm, simply covering the QDs with a GaAs layer only a few nanometres thicker than the QD height may not necessarily provide an optimal strain environment for subsequent thin capping.
For this reason, in our samples, 5\,nm of GaAs was initially deposited over the QDs at 485$^\circ$C, after which the growth was interrupted and the substrate temperature was increased to 580$^\circ$C under an As$_{4}$ overpressure. Thermal treatment of partially capped InAs/GaAs QDs is known to induce QD reshaping and material redistribution \cite{Ref3}. Although the temperature employed here is lower than that typically used for a conventional In-flush process, we expect that the thermal treatment may promote lateral compositional intermixing between the InAs QDs and the surrounding GaAs. Such intermixing would result in a more gradual lateral compositional profile and an associated redistribution of the strain field around the QDs.

We therefore speculate that this lateral compositional intermixing and strain redistribution may be among the factors allowing the QDs to retain effective three-dimensional carrier confinement after the growth of only an additional 5 or 15\,nm of GaAs. The resulting QDs would not simply correspond to as-grown InAs QDs buried beneath a thin GaAs layer, but rather to thermally modified In(Ga)As/GaAs nanostructures with a reduced and more gradually distributed lattice mismatch. This may reduce the sensitivity of the confined carriers to the nearby surface and could partly explain why the excitonic lifetimes remain close to 1\,ns even for the thin capping layers investigated here.

In addition to the compositional intermixing induced by the thermal treatment, the small thickness of the GaAs capping layer itself may influence the strain distribution around the QDs. In conventional structures with a thick GaAs cap, the InAs QDs are embedded in an effectively bulk-like GaAs matrix and are therefore strongly constrained by the surrounding material. In the present structures, however, the proximity of the surface may allow partial relaxation of the thin GaAs layer above the QDs. 
Such an effective strain-relaxation environment may have an effect analogous to that observed when strain-reducing layers, such as InGaAs, are introduced above InAs QDs, where the reduced strain is known to shift the QD emission towards longer wavelengths \cite{Ref4}. We therefore expect that the thin GaAs capping layer may itself act, to some extent, as an effective strain-relaxation layer because of its thin thickness. This effect could contribute to the relatively long-wavelength emission observed from the thin-capped QDs, despite the compositional intermixing induced by the thermal treatment, which would otherwise be expected to produce a blueshift.

In summary, our steady-state and time-resolved photoluminescence measurements show that the QDs under study retain good optical properties at close proximity to the surface, a significant advancement compared to previous reports \cite{wang_optical_2004, chellu_gaas_2021}. 
Given that the coupling between quantum light sources and plasmonic structures is a near-field dipole-dipole coupling, which requires nanometric separation between the emitter and the structure \cite{bar-david_enhanced_2023,indukuri_ultrasmall_2019,chikkaraddy_single-molecule_2016,baumberg_extreme_2019,stern_transmission_2012}, effective coupling requires QDs to be in close proximity to fabricated plasmonic structures. 
Our results, that show that epitaxial QDs can retain good optical properties in proximity to a surface, are significant for a wide array of photonic applications, and particularly for the integration with plasmonic structures, that can increase the brightness and directionality of quantum emitters \cite{our_rings1, our_rings2, our_rings3, our_rings4, livneh_highly_2011,bar-david_enhanced_2023}, therefore opening the path to quantum plasmonic devices with solid-state epitaxial quantum emitters \cite{q_plasmon}. 

\section*{Acknowledgements}
We would like to thank Jeremy Baumberg and Edmund Clarke for useful discussions, and the Engineering and Physical Sciences Research Council (grants EP/X040526/1, EP/Z533178/1) and The Leverhulme Trust (grant RPG-2021-305) for financial support.


\end{document}